\documentclass[showpacs,preprintnumbers,amsmath,aps,amssymb]{revtex4}
\usepackage{graphicx}% Include figure files
\usepackage{dcolumn}% Align table columns on decimal point
\usepackage{bm}% bold math
\begin{document}
\title{Quintom potentials from quantum cosmology using the FRW cosmological model}
\author{J. Socorro $^{1,2}$}
\email{socorro@fisica.ugto.mx}
\author{Priscila Romero$^3$}
\email{kamadi_shika@live.com.mx}
\author{Luis O. Pimentel $^2$}
\email{lopr@xanum.uam.mx}
\author{M. Aguero $^3$}
\email{makxim@gmail.com} \affiliation{$^1$Departamento de
F\'{\i}sica, DCeI, Universidad de Guanajuato-Campus Le\'on,
 C.P. 37150, Le\'on, Guanajuato, M\'exico\\
  $^2$ Departamento de F\'isica, Universidad Aut\'onoma Metropolitana, Apartado Postal 55-534,
C.P. 09340 M\'exico, DF, M\'exico\\
  $^3$ Facultad de Ciencias de la Universidad Aut\'onoma del Estado de M\'exico, Instituto Literario
No. 100, C.P. 50000, Toluca, Edo de Mex, Mexico
  }%

\date{\today}

\begin{abstract}
We construct the quintom potential of dark energy models in the
framework of spatially flat
  Friedmann-Robertson Walker universe in the inflationary epoch, using the Bohm like approach, known as
 amplitude-real-phase. We find some potentials for which the wave
 function of the universe is found analytically and we have obtained the classical trajectories in the
 inflation era.
\end{abstract}

\maketitle
\section{Introduction}
At the present time, there are some efforts to explain the
observations for the accelerated expansion of the universe,  based
on the dynamics of a scalar (quintessence) or multiscalar field
(quintom) as models for  dark energy, (see the review
\citep{copeland,feng,cai}). The properties of the quintom models
have been studied from different points of view. Among them, the
phase space studies, using the dynamical systems tools, are very
useful in order to analyze the qualitative and asymptotic behavior
of the model
    \citep{guo1,feng0,sadjadi,guo2,feng2,zhao,cai2,cai3,alimo,lazkoz1,cai4,cai5,zhang,sadeghi1,nozari,setare1,setare0,sadeghi2,qiu,saridakis,amani,fara}).
    In the present work we want to
investigate  the case of quintom cosmology constructed by using both
quintessence ($\sigma$) and phantom ($\phi$) fields, maintaining a
nonspecific potential  $\rm V(\phi,\sigma)$ from the begining. In
the literature one special class of potential used is the sum of the
exponential potentials for each field
\citep{guo1,sadjadi,zong,zhao,chimento}.
 There are other works where other type of
potentials are analyzed
\cite{copeland,zhao,setare0,cai5,adak,setare1,saridakis,amani}.

We claim that the analysis of general potentials using dynamical
systems was made considering particular structures of them,  in
other words, how can we introduce this mathematical structure within
a physical context? We can partially answer this question, when the
amplitude-real-phase formalism is introduced, in the sense that,
some potentials can be constructed \citep{wssa,sodo,bohm}.

This work is arranged as follows. In section \ref{model} we present
the corresponding Einstein-Klein-Gordon equation for the quintom
model. Also we introduced the hamiltonian apparatus which is applied
to FRW cosmological model in order to construct a master equation
for barotropic perfect fluid and cosmological constant. Furthermore,
we present the classical equations for the flat FRW, which are
presented in section \ref{hamil} for particular scalar potentials.
In subsection \ref{qap} we present the quantum scheme, where we will
use the amplitude-real-phase ansatz in order to solve the WDW
equation. This method was used by Moncrief and Ryan \cite{mo-ry} to
obtain exact solutions for the Bianchi type IX anisotropic
cosmological model with a wavefunction of the form $\rm \Psi=W
e^{-S}$. This form of the wavefunction was motivated by the work of
Kodama \cite{kodama1,kodama2}. Our treatment is applied to build the
mathematical structure  of quintom scalar potentials which allow us
obtain exact solutions together with an arbitrary factorization of
the WDW equation. In this work we have obtain the wave function of
the universe   in order to find the quantum potentials, which is a
more important matter in order to find the classical trajectories,
      which is shown in section \ref{hamil}, that is devoted to obtain the classical solutions for particular scalar
       potentials, also we show graphically the classical trajectory in the configuration space $\rm (\Omega,\varphi)$
       projected from its quantum counterpart. In this same
       section, we present the time dependence for the $\Omega$, and quintom scalar fields ($\varphi\, ,
       \varsigma$) for the potentials treated in this section. In
       section \ref{remarks} is devoted to final remarks of this
       work.

\section{The Quintom model \label{model}}
We begin with the construction of the quintom cosmological theory,
which requires the simultaneous consideration of two fields, namely,
a canonical one $\sigma$, and  phantom one $\phi$ (see equation
(\ref{densidad}) below and reference \cite{setare0}). The lagrangian
density of
 this theory with two fields, the
cosmological term,  and  matter  contribution,  is

\begin{equation}
\rm {\cal L}=\sqrt{-g} \left( R-2\Lambda
-\frac{1}{2}g^{\mu\nu}\nabla_\mu \phi \nabla_\nu \phi
+\frac{1}{2}g^{\mu\nu}\nabla_\mu \sigma \nabla_\nu \sigma  -
V(\phi,\sigma)\right)+ {\cal L}_{matter}, \label{lagra}
\end{equation}
the corresponding field equations for perfect fluid matter content
are
\begin{eqnarray}
  \rm G_{\alpha \beta}+ g_{\alpha \beta} \Lambda &=&\rm -\frac{1}{2}\left(\nabla_\alpha \phi \nabla_\beta \phi -\frac{1}{2}g_{\alpha \beta} g^{\mu \nu}
\nabla_\mu \phi \nabla_\nu \phi \right) +
\frac{1}{2}\left(\nabla_\alpha \sigma \nabla_\beta \sigma
-\frac{1}{2}g_{\alpha \beta} g^{\mu \nu}
\nabla_\mu \sigma \nabla_\nu \sigma \right) \nonumber \\
&& \rm -\frac{1}{2}g_{\alpha \beta} \, V(\phi,\sigma) -8\pi G T_{\alpha \beta}, \label{munu}\\
\rm g^{\mu\nu} \phi_{,\mu\nu} -g^{\alpha \beta} \Gamma^\nu_{\alpha
\beta} \nabla_\nu \phi + \frac{\partial V}{\partial \phi}&=&\rm
0,\qquad
\Leftrightarrow \quad \Box \phi +\frac{\partial V}{\partial \phi}=0 \nonumber\\
\rm g^{\mu\nu} \sigma_{,\mu\nu} -g^{\alpha \beta} \Gamma^\nu_{\alpha
\beta} \nabla_\nu \sigma - \frac{\partial V}{\partial \sigma}&=&\rm
0,\qquad
\Leftrightarrow \quad \Box \sigma -\frac{\partial V}{\partial \sigma}=0,\nonumber\\
 \rm T^{\mu\nu}_{\,\,;\mu}&=&\rm 0,\quad with \quad T_{\mu\nu}= Pg_{\mu\nu}+(P+\rho)u_\mu u_\nu, \label{energy-momentum}
\end{eqnarray}
here $\rho$ is the energy density, P the pressure, and   $\rm u_{\mu
}$ the velocity of the fluid, satisfying  that $\rm u_\mu u^\mu=-1$.

%%%%%%%%%%%%%%%%%%%%%%%%%%%%%%%%

Let us recall here the
canonical formulation in the ADM formalism for flat FRW cosmological
model. The metric has the form
\begin{equation}
\rm ds^2= -N(t)dt^2 + e^{2\Omega(t)}\left[
dr^2 + r^2\left(d\theta^2 +\sin^2\theta
d\phi^2\right)\right],\label{metricI}
\end{equation}
where the scale factor is defined as $\rm A=e^{\Omega(t)}$, where
$\rm \Omega \epsilon\,(-\infty,\infty)$.

 The lagrangian density (\ref{lagra}) for  FRW cosmological
 model is written as
 \begin{equation}
\rm {\cal L}_{FRW} = \rm e^{3\Omega}\left[6 \frac{\dot \Omega^2}{N}
+ 6 \frac{\dot \varphi^2}{N}-6 \frac{\dot \varsigma^2}{N}+ N\left(-
V(\varphi,\varsigma)+2\Lambda+16\pi G   \rho   \right)  \right],
\label{lagra-i}
\end{equation}
 where the overdot denotes time derivative,
the fields were re-scaled as $\phi=\sqrt{12}\varphi,
\sigma=\sqrt{12}\varsigma$ for simplicity in the calculations.

The momenta are define in the usual way, $\rm \Pi_{q^i}=\rm
\frac{\partial {\cal L} }{\partial \dot q^i}$, where $\rm q^i=(A,
\varphi,\varsigma)$ are the fields coordinates of the model.
\begin{eqnarray}
\rm \Pi_\Omega  &=&\rm \frac{\partial {\cal L}}{\partial \dot
\Omega} = \frac{12e^{3\Omega} \dot\Omega}{N}, \qquad \to
\dot \Omega = \frac{N\Pi_\Omega}{12}e^{-3\Omega} \nonumber\\
\rm \Pi_{\varphi}  & =&\rm \frac{\partial {\cal L}}{\partial \dot
\varphi} =12\frac{e^{3\Omega} \dot\varphi}{N}, \qquad  \to
\dot{{\varphi}}  = +\frac{N\Pi_\varphi}{12}  e^{-3\Omega}\nonumber\\
\rm \Pi_{\varsigma} &=&\rm \frac{\partial {\cal L}}{\partial \dot
\varsigma} =-12\frac{e^{3\Omega} \dot\varsigma}{N}, \qquad  \to
\dot{{\varsigma}}  = -\frac{N\Pi_\varsigma}{12} e^{-3\Omega}.
\label{momenta}
\end{eqnarray}
Writing (\ref{lagra-i}) in the canonical form,
 $\rm {\cal L}_{canonical}=\Pi_q \dot q -N{\cal H}$  and substituting the energy density of the barotropic fluid $\rm P=\gamma \rho$,
 which was found using the covariant derivative of the energy-momentum tensor (\ref{energy-momentum}),
 $
\rm \rho =M_\gamma e^{-3 \left( 1+\gamma \right)\Omega},$ where
$M_\gamma$ is an integration constant.
  The corresponding
 Hamiltonian density  ${\cal H}$
 considering in this work the inflationary
epoch, that corresponds to the value $\gamma=-1$, becomes
\begin{equation}
\rm {\cal H}_{FRW}
=\frac{e^{-3\Omega}}{24}\left[\Pi_\Omega^2-\Pi_\varsigma^2+\Pi_\varphi^2
    +e^{6\Omega}\left\{24V(\varphi,\varsigma)-\lambda_{eff}\right\}
\right], \label{hami-i}
\end{equation}
 where $\rm \lambda_{eff}=48(\Lambda+8\pi G M_{_{-1}})$.
%\subsection{Classical field equations for the flat FRW}

The Einstein field equations (\ref{munu},\ref{energy-momentum}) for
the flat FRW cosmological model with barotropic state equation, are
\begin{eqnarray}
\rm 3\frac{\dot \Omega^2}{N^2}&=&\rm 8\pi G \rho-3\frac{\dot
\varphi^2}{N^2}+3\frac{\dot \varsigma^2}{N^2} +
\Lambda+\frac{V(\varphi,\varsigma)}{2}, \label{00} \\
\rm 2\frac{\ddot \Omega}{N^2}+3\frac{\dot \Omega^2}{N^2}
-2\frac{\dot \Omega \dot N}{N^3}&=&\rm-8\pi G P+ 3\frac{\dot
\varphi^2}{N^2}-3\frac{\dot \varsigma^2}{N^2}+\Lambda
+\frac{V(\varphi,\varsigma)}{2},
\label{11} \\
\rm -3\frac{\dot \Omega \dot \varsigma}{N^2}+\frac{\dot N \dot
\varsigma}{N^3}-\frac{\ddot \varsigma}{N^2}+\frac{\partial
V(\varsigma,\varphi)}{\partial \varsigma}&=&0,
\label{da-sigma}\\
\rm -3\frac{\dot \Omega \dot \varphi}{N^2}+\frac{\dot N \dot
\varphi}{N^3}-\frac{\ddot \varphi}{N^2}-\frac{\partial
V(\varsigma,\varphi)}{\partial \varphi}&=&0, \label{da-phi}
\end{eqnarray}
which can be written as
\begin{eqnarray}
\rm 8\pi G \rho + \Lambda + \frac{1}{2}\left(-6\varphi^{\prime
2}+6\varsigma^{\prime 2}+V(\varphi,\varsigma)  \right) &=& \rm 3H^2,
\label{densidad}\\
\rm 8\pi G P- \Lambda+ \frac{1}{2}\left(-6\varphi^{\prime
2}+6\varsigma^{\prime 2}-V(\varphi,\varsigma)
\right)=-2\Omega^{\prime\prime}-3H^2 &=& \rm 3H^2q-H^2,
\label{presion}\\
\rm -3 \Omega^\prime \varsigma^\prime-
\varsigma^{\prime\prime}+\frac{\partial
V(\varsigma,\varphi)}{\partial \varsigma}&=&0,
\label{da-varsigma}\\
\rm -3 \Omega^\prime \varphi^\prime-
\varphi^{\prime\prime}-\frac{\partial V(\varsigma,\varphi)}{\partial
\varphi}&=&0, \label{da-varphi}
\end{eqnarray}
where the Hubble parameter is define as $\rm H=\frac{\dot
A}{A}=\dot\Omega$ and the deceleration parameter $q=-\frac{A \ddot
A}{\dot A^2}$.  We have done the time transformation $\rm
\frac{d}{d\tau}=\frac{d}{N dt}= \prime$.

Adding  (\ref{densidad})  and (\ref{presion}) we obtain
\begin{equation}
\rm -\Omega^{\prime\prime}=4\pi G\left[\rho +\rho_\varphi+
\rho_\varsigma +P + P_\varphi + P_\varsigma \right],
\end{equation}
where
\begin{eqnarray}
\rm P_\varphi&=&\rm\frac{1}{16\pi G}\left( -6\varphi^{\prime
2}-V(\varphi,\varsigma)|_{\varsigma}\right),
\qquad P_\varsigma=\frac{1}{16\pi G} \left(6\varsigma^{\prime 2}-V(\varphi,\varsigma)|_{\varphi}\right),\nonumber\\
\rm \rho_\varphi&=&\rm\frac{1}{16\pi G} \left( -6\varphi^{\prime
2}+V(\varphi,\varsigma)|_{\varsigma}\right), \qquad
\rho_\varsigma=\frac{1}{16\pi G} \left(6\varsigma^{\prime
2}+V(\varphi,\varsigma)|_{\varphi}\right),\nonumber
\end{eqnarray}
which are useful when we study the behavior of dynamical systems.
Additionally we can introduce the total quintom energy density and
pressure as:
\begin{equation}
    \rho_{_{DE}}=\rho_{\varsigma}+\rho_{\varphi},\qquad P_{_{DE}}=P_{\varsigma}+P_\varphi, \qquad
    P_{_{DE}}=\omega_{_{DE}}\rho_{_{DE}}
\end{equation}
where
\begin{equation}\rm
    \omega_{_{DE}}=\frac{6\varsigma^{\prime2}-6\varphi^{\prime 2}-V(\varsigma,\varphi)}{6\varsigma^{\prime 2}-6\varphi^{\prime 2}+V(\varsigma,\varphi)}=
    -1 -\frac{12\left(\varphi^{\prime 2}-\varsigma^{\prime2} \right)}{V(\varsigma,\varphi)-6\left(\varphi^{\prime 2}-\varsigma^{\prime2} \right)}.
    \label{baro}
\end{equation}

For phenomenological analysis, (\ref{baro}) will serve us as a test
for the viable  class of quintom potentials whose equation of state
can cross the cosmological constant barrier which
$\omega_{_{DE}}=-1$ is mildly favored of observations
\cite{melchiorri}. For instance, when $\rm V(\varsigma,\varphi)>0$
and $\rm V(\varsigma,\varphi)>6\varphi^{\prime
2}-6\varsigma^{\prime2}$, then the kinetic terms satisfies $\rm
\varsigma^{\prime 2}<\varphi^{\prime2}$, so, the barotropic
parameter $\rm \omega_{_{DE}}<-1$. In this case, the intensity of
the phantom field $\rm \varphi$ always is larger than the intensity
of the quintessence field $\rm \varsigma$. In other case, we have
that $\rm \omega_{_{DE}}> -1$. The particular value equivalent to
cosmological constant is when the kinetic terms are null or equal in
both fields.

For constant potential, equations
(\ref{da-varsigma},\ref{da-varphi}) can be solved in terms of the
$\Omega$ function,
\begin{eqnarray}
\rm \varsigma^\prime&=&\rm \varsigma_0 e^{-3 \Omega}, \qquad
\varsigma(\tau)=\varsigma_0\int e^{-3 \Omega(\tau) } d\tau
+\varsigma_1,\label{varvar}\\
 \rm \varphi^\prime&=&\rm \varphi_0 e^{-3
\Omega},\qquad \varphi(\tau)=\varphi_0\int e^{-3 \Omega(\tau) }
d\tau +\varphi_1,\label{phiphi}
\end{eqnarray}
We obtain in implicit solution for $\Omega$ in a quadrature
\begin{equation}
\rm \int \frac{d\Omega}{\sqrt{F(\Omega)}}=\Delta \tau, \label{maese}
\end{equation}
where  $\rm F(\Omega)$ is given by
$$\rm F(\Omega)=\frac{8\pi G M_\gamma}{3}e^{-3(1+\gamma)\Omega}+\left(-\varphi_0^2+\varsigma_0^2 \right)e^{-6\Delta \Omega}
+\frac{\Lambda_{eff}}{3},$$ with $\rm \Lambda_{eff}=\Lambda+V_0/2.$
For particular values of the parameter $\gamma$, has analytical
solution.

 For one scalar field, with constant potential, the formalism is like the one formulated by
S\'aez and Ballester in 1986,  because both field are equivalent,
see equations (\ref{da-varsigma},\ref{da-varphi})
\cite{saez-ballester}.
  This formalism was studied by one of
the author and collaborators, in the FRW and Bianchi type Class A
cosmological models, \cite{soco1,soco2,soco3}.

 \section{quantum cosmology\label{qap}}
The Wheeler-DeWitt equation for this model is obtained introducing
the representation (we choose $\hbar=1$)
 $\rm \Pi_{q^\mu}=-i \partial_{q^\mu}$ into (\ref {hami-i}). The factor $\rm e^{-3\Omega}$ can be ordered
 with the momenta $\rm \hat \Pi_\Omega$ in many ways. Hartle and
Hawking \citep{HH} have suggested a semi-general factor ordering for
$\rm e^{-3\Omega} \hat \Pi^2_\Omega$ as
\begin{eqnarray}
\rm - e^{-(3- Q)\Omega}\, \partial_\Omega e^{-Q\Omega}
\partial_\Omega&=&\rm - e^{-3\Omega}\, \partial^2_\Omega + Q\,
e^{-3\Omega} \partial_\Omega, \label {hh}
\end{eqnarray}
where  Q is any real constant that measure the ambiguity in the
factor ordering in the variable $\Omega$. In the following we will
assume this factor ordering for the Wheeler-DeWitt equation, which
becomes
\begin{equation}
\rm \Box \Psi+ Q\frac{\partial \Psi}{\partial \Omega}+e^{6\Omega}
U(\Omega,\varphi,\varsigma,\lambda_{eff})\Psi=0, \label{wdwmod}
\end{equation}
where $\rm \Box=-\frac{\partial^2}{\partial
\Omega^2}-\frac{\partial^2}{\partial \varphi^2}
+\frac{\partial^2}{\partial \varsigma^2}$ is the d'Alambertian in
the field coordinates $q^\mu=(\Omega,\varsigma,\varphi)$ and $\rm
U=24V(\varphi,\varsigma)-\lambda_{eff}$.

\subsection{Solving the WDW equation in amplitude-real-phase approach}
Some time ago, Moncrief and Ryan tried successfully
amplitude-real-phase ansatz in order to solve the WDW equation  to
obtain exact solutions for the Bianchi type IX anisotropic
cosmological model. This ansatz was motivated by the remarks of
Kodama \cite{kodama1,kodama2} on the relation between the ADM
wavefunction and the Ashtekar wavefunction. This ansatz is as
follows
\begin{equation}
\rm \Psi(\ell^\mu) = W(\ell^\mu) e^{- S(\ell^\mu)}, \label{ans}
\end{equation}
where $S(\ell^\mu)$ is known as the superpotential function, W is
like  probability amplitude in the Bohm formalism  \citep{bohm}.

Then, the WDW equation (\ref{wdwmod}) is written as
\begin{equation}
 \left\{{\Box \, W}  -
Q \frac{\partial W}{\partial \Omega}\right\}- \left\{ W \left({\Box
\, S}-Q  \frac{\partial S}{\partial \Omega}\right) + 2 {\nabla
W}\cdot {\nabla S}\right\} + \left\{
 W \left[ \left(\nabla S\right)^2 - {\cal U}\right]\right\} = 0,
\label {mod}
\end{equation}
with $\rm \Box = G^{\mu \nu}\frac{\partial^2}{\partial \ell^\mu
\partial \ell^\nu}$, $\rm {\nabla \, W}\cdot {\nabla \, \Phi}=G^{\mu
\nu} \frac{\partial W}{\partial \ell^\mu}\frac{\partial
\Phi}{\partial \ell^\nu}$, $\rm (\nabla)^2= G^{\mu
\nu}\frac{\partial }{\partial \ell^\mu}\frac{\partial }{\partial
\ell^\nu}= +(\frac{\partial}{\partial \varphi})^2
+(\frac{\partial}{\partial \Omega})^2 - (\frac{\partial}{\partial
\varsigma})^2$, where $\rm G^{\mu \nu}= diag(-1,1,1)$, and $\rm
{\cal U}=e^{6\Omega}U(\varphi,\varsigma,\lambda_{eff})$ is the
potential term.

The last equation is a difficult  one to solve, in view of which we
propose to factorize as below and set each factor to zero, obtaining
simpler set of equations,
\begin{subequations}
\label{WDWa}
\begin{eqnarray}
(\nabla S)^2 - {\cal U} &=& 0, \label{hj} \\
       W \left( \Box S - Q \frac{\partial S}{\partial \Omega}
  \right) + 2 \nabla \, W \cdot \nabla \, S &=& 0 \, ,
  \label{wdwho}\,\\
  \Box \, W - Q \frac{\partial W}{\partial \Omega} & = & 0, \label{cons}
\end{eqnarray}
\end{subequations}

We follow the approach presented in the references \citep{wssa,sodo}
for solving these set of partial differential equations in order to
obtain particular solutions to de WDW equation. We solve initially
the Einstein-Hamilton-Jacobi (EHJ) equation (\ref{hj}), next we
introduce the superpotential function S into equation (\ref{wdwho})
in order to solve for the W function, and finally, these solutions
must satisfy the quantum potential equation (\ref{cons}), that
appears as a quantum constraint in our approach.

\section{Mathematical structure of quintom potential}
To solve  Hamilton-Jacobi equation (\ref{hj})
$$\rm \left(\frac{\partial S}{\partial \varphi} \right)^2+\left(\frac{\partial S}{\partial \Omega} \right)^2
-\left(\frac{\partial S}{\partial \varsigma} \right)^2=e^{6\Omega}
U(\varphi,\varsigma,\lambda_{eff})$$
 we propose that the superpotential function has the following
 form
\begin{equation}
\rm S=e^{3\Omega} g(\varphi) h(\varsigma), \label{sp}
\end{equation}
 and substituting into (\ref{hj}),
\begin{equation}
  \rm  e^{6\Omega}\left[h^2\left(\frac{dg}{d\varphi}
\right)^2-g^2\left(\frac{dh}{d\varsigma} \right)^2+9g^2 h^2-
U(\varphi,\varsigma,\lambda_{eff}) \right]=0, \label{ecuacion}
\end{equation}
in order to solve by separation of variables, this equation imply
the foolowing structure for the potential
\begin{equation} \rm U=g^2 h^2 \left[c_1 G(g) + c_0 H(h)\right], \label{potenti-u}
\end{equation}
where $\rm g(\varphi)$, $\rm h(\varsigma)$, $\rm G(g)$ and $\rm
H(h)$ are generic functions on their arguments, which will be
determined under this process, and $\rm c_0$, $\rm c_1$ are
constants. Then, by method of separation variables we find the
following master equations for the quintom fields
\begin{subequations}
\label{fields}
\begin{eqnarray}
\rm d\varphi&=&\rm \pm \frac{dg}{g\sqrt{p^2+ c_1 G}}, \qquad with \quad p^2=\nu^2-\frac{9}{2},\label{varsigma}\\
\rm d\varsigma&=&\rm \pm \frac{dh}{h\sqrt{\ell^2- c_0 H}}, \qquad
with \quad \ell^2=\nu^2+\frac{9}{2},\label{varphi}
\end{eqnarray}
\end{subequations}
where $\nu$ is a constant of separation of variables.

For  particular choices of functions $\rm G$ and $\rm H$  we can
solve for the $\rm g(\varphi)$ and $\rm h(\varsigma)$ functions, and
then use them to obtain the potential term U from (\ref{potenti-u}).
 Some
examples are shown in the tables \ref{t:solutions} and
\ref{potentials}, thereby, the superpotential $\rm
S(\Omega,\varphi)$ is known, and the possible quintom potentials are
shown in table \ref{potentials}.

\begin{center}
    \begin{table}[h]
            \begin{tabular}{|c|c|c|c|}
            \hline
            $\rm  H(h)$ & $\rm h(\varsigma)$ &$\rm G(g)$& $\rm g(\varphi)$ \\ \hline
            $ 0 $ & $\rm h_0 e^{\pm \ell  \Delta \varsigma}$&  0& $\rm g_0 e^{\pm p  \Delta \varphi}$  \\ \hline
            $\rm H_0$ & $\rm h_0 e^{\pm \sqrt{\ell^2-c_0 \, H_0} \Delta \varsigma}$& $\rm  G_0$ &$\rm g_0 e^{\pm \sqrt{p^2+c_1 \,G_0} \Delta \varphi}$\\ \hline
            $\rm  H_0 h^{-2} $ &$\rm \frac{\sqrt{c_0H_0}}{\ell} cosh\left[\ell \Delta \varsigma \right]  $&  $\rm  G_0 g^{-2}  $
            & $\rm \frac{\sqrt{c_1 \, G_0}}{p} sinh \left[p \Delta \varphi \right] $ \\ \hline
            $\rm H_0 h^{-n}$ ($\rm n \neq 2$) &  $\rm \left[\frac{c_0H_0}{\ell^2}\, cosh^2\left(\frac{n \ell \Delta \varsigma }{2} \right)\right]^{1/n}$ &
            $\rm G_0 g^{-n}$ ($\rm n \neq 2$)&  $\rm \left[\frac{c_1G_0}{p^2}\, sinh^2\left(\frac{n p\Delta \varphi}{2}\right)\right]^{1/n}$\\\hline
            $\rm  H_0\, \ln h$ & $\rm e^{u(\varsigma)}$, & $\rm  G_0\, \ln g$& $\rm e^{v(\varphi)}$\\
            & $\rm u(\varsigma)=\frac{\ell^2-\left(\frac{c_0H_0}{2}\Delta \varsigma \right)^2}{c_0H_0}$&
            &$\rm v(\varphi)=\frac{-p^2+\left(\frac{c_1G_0}{2}\Delta \varphi \right)^2}{c_1G_0}$\\\hline
            $\rm H_0  (\ln h)^2$ & $\rm e^{r(\varsigma)} $ & $\rm G_0  (\ln g)^2$ &$\rm e^{\omega(\varphi)} $\\
            & $\rm r(\varsigma)=\frac{\ell}{\sqrt{c_0H_0}}sin\left(\sqrt{c_0H_0}\,  \Delta \varsigma \right) $&
            &$\rm \omega(\varphi)=\frac{p}{\sqrt{c_1G_0}}sinh\left(\sqrt{c_1G_0}\,  \Delta \varphi \right)$ \\  \hline
            \end{tabular}
             \caption{ \label{t:solutions} \emph{Some exact solutions to eqs.
(\ref{varsigma},\ref{varphi}), where n is any real number, $\rm G_0$
and $\rm H_0$ are an arbitrary constants.}}
        \end{table}
    \end{center}

    \begin{center}
        \begin{table}[h]
             \begin{tabular}{|c|c|}
            \hline
            $\rm  U(\varphi,\varsigma)$ & Relation between all constants  \\ \hline
            0 &  $\rm \ell^2(s-p^2-3k-9)^2-p^2(s-\ell^2)^2 +\ell^2 p^2(k^2-Q^2)=0$  \\ \hline
            $\rm U_0 e^{\pm 2[\sqrt{\ell^2-c_0H_0}\Delta \varsigma + \sqrt{p^2+c_1G_0}\Delta \varphi]}$ &
             $\rm(\ell^2-c_0H_0) (s-p^2-3k-9-c_1G_0)^2-( p^2+c_1G_0)(s-\ell^2+c_0H_0)^2 $\\
                 & $\rm  +(\ell^2-c_0H_0)( p^2+c_1G_0)(k^2-Q^2)=0$\\             \hline
            $\rm U_0 sinh^2(p\Delta \varphi)+U_1\, cosh^2(\ell \Delta \varsigma)$&$\rm k(k-6)=Q^2,\,\,6k(9+p^2)+9Q^2-p^4+(\ell^2-9)^2=0,\,\,
             \,s=\ell^2$,\\ \hline
            {\small{$\rm b_0H_0 \left[\frac{c_1G_0}{p^2} sinh^2\left(\frac{n p\Delta \varphi}{2}\right)\right]^{\frac{2}{n}}
            \left[\frac{c_0H_0}{\ell^2} cosh^2\left(\frac{n}{2}\ell\Delta \varsigma\right)\right]^{\frac{2-n}{n}} +$}} &
            $\rm  \mbox{quantum constraint is not satisfied} $   \\
            {\small{$\rm a_0G_0 \left[\frac{c_1G_0}{p^2} sinh^2\left(\frac{n p\Delta \varphi}{2}
            \right)\right]^{\frac{2-n}{n}}
            \left[\frac{c_0H_0}{\ell^2} cosh^2\left(\frac{n}{2}\ell\Delta \varsigma\right)\right]^{\frac{2}{n}} $}} &\\ \hline
            $\rm e^{2u(\varsigma)+2v(\varphi)}\left[b_0H_0 u(\varsigma)+
            a_0G_0\,v(\varphi) \right]$ &$\rm  \mbox{quantum constraint is not satisfied} $\\\hline
                        $\rm e^{2r(\varsigma)+2\omega(\varphi)}\left[b_0H_0 r^2 +a_0G_0 \omega^2 \right]$&  $
                        \mbox{quantum constraint is not satisfied} $ \\ \hline
                      \end{tabular}
                      \caption{ \label{potentials} \emph{The corresponding quintom potentials
that emerge from quantum cosmology in direct relation with the table
(\ref{t:solutions}). Also we present the relation between all
constant that satisfy the eqn. (\ref{cons}). We can see that the
quantum constraint restrict the general potential of fifth line to
remain in the state of n=2. The sixth and seventh lines indicate
that these potentials are not allowed. }}
        \end{table}
    \end{center}

To solve (\ref{wdwho}) we assume  the separation variables for the W
function
\begin{equation} W=e^{\left[
\eta(\Omega)+\xi(\varphi)+\lambda(\varsigma) \right]}, \label{ww}
\end{equation}
and introducing the corresponding superpotential function S
(\ref{sp}) into the equation (\ref{wdwho}), it follows the equation
\begin{equation}
\rm \frac{1}{g}
\frac{d^2g}{d\varphi^2}+\frac{2}{g}\frac{dg}{d\varphi}
\frac{d\xi}{d\varphi}+9-\frac{1}{h}\frac{d^2h}{d\varsigma^2}-\frac{2}{h}\frac{d\lambda}{d\varsigma}\frac{dh}{d\varsigma}+6\frac{d\eta}{d\Omega}-3Q=0,
\label{solve}
\end{equation}
and using the method of separation of variables,
  we arrive to a set of ordinary differential equations for
the functions $\eta(\Omega)$, $\rm \xi(\varphi)$ and
$\lambda(\varsigma)$ (however, this decomposition is not unique,
because it depends on the way as we put the constants in the
equations).
\begin{eqnarray}
\rm 2\frac{d\eta}{d\Omega}-Q&=&\rm k, \\
\rm \frac{d^2g}{d\varphi^2} +2\frac{dg}{d\varphi} \frac {d\xi}{d\varphi}&=& \rm [s-3(k+3)]g,\\
\rm \frac{d^2h}{d\varsigma^2}+2\frac{dh}{d\varsigma}
\frac{d\lambda}{d\varsigma}&=& \rm sh,
\end{eqnarray}
whose solutions in the generic fields g and h are (given in table
(\ref{t:solutions}))
\begin{eqnarray}
\rm \eta(\Omega)&=& \rm \frac{Q+k}{2} \Omega, \nonumber\\
\rm \lambda(\varsigma)&=&\rm \frac{s}{2}\int
\frac{d\varsigma}{\partial_\varsigma(ln h)}-\frac{1}{2}\int
\frac{\frac{d^2h}{d\varsigma^2}}{\partial_\varsigma h}d\varsigma,
\nonumber\\
\rm \xi(\varphi)&=& \rm \left(\frac{s}{2} -\frac{3k}{2}
-\frac{9}{2}\right)\int \frac{d\varphi}{\partial_\varphi(ln g)}
-\frac{1}{2}\int \frac{\frac{d^2g}{d\varphi^2}}{\partial_\varphi
g}d\varphi, \nonumber
\end{eqnarray}
then the solution for the function W is
\begin{equation}
\rm W=e^{ \frac{s}{2}\int
\left(\frac{d\varsigma}{\partial_\varsigma(ln
h)}+\frac{d\varphi}{\partial_\varphi(ln g)} \right)} e^{-\frac{1}{2}
\int \left(\frac{\frac{d^2h}{d\varsigma^2}}{\partial_\varsigma
h}d\varsigma+ \frac{\frac{d^2g}{d\varphi^2}}{\partial_\varphi
g}d\varphi \right)} e^{\frac{k}{2}\left(\Omega-3\int
\frac{d\varphi}{\partial_\varphi(ln g)}\right)}
e^{\frac{1}{2}\left(Q\Omega-9\int
\frac{d\varphi}{\partial_\varphi(ln g)}\right)}. \label{ww2}
\end{equation}

In a similar way, the constraint (\ref{cons}) can be written as
\begin{equation}
 \rm \partial^2_{\varphi} \xi + \left(\partial_{\varphi} \xi \right)^2-\partial^2_{\varsigma} \lambda
 - \left( \partial_{\varsigma} \lambda \right)^2+\frac{k^2-Q^2}{4} = 0 \, , \label{wdwho2}
\end{equation}
or in other words (here $\rm \mu=s-3(3+\kappa)$)
\begin{equation}
 \rm
2\frac{\partial^3_\varsigma h}{\partial_\varsigma
h}-2\frac{\partial^3_\varphi g}{\partial_\varphi
g}+4sh\frac{\partial^2_\varsigma h}{(\partial_\varsigma h)^2}-4\mu
g\frac{\partial^2_\varphi g}{(\partial_\varphi
g)^2}-3\frac{(\partial^2_\varsigma h)^2}{(\partial_\varsigma
h)^2}+3\frac{(\partial^2_\varphi g)^2}{(\partial_\varphi
g)^2}-\frac{s^2 h^2}{(\partial_\varsigma h)^2}+\frac{\mu^2
g^2}{(\partial_\varphi g)^2}-2s+2\mu+k^2-Q^2=0. \nonumber
\end{equation}

When we use the different cases presented in the table
(\ref{t:solutions}), the  relations between all constants were
found, which we presented in the same table \ref{potentials} with
the quintom potentials. So, the quantum solutions for each potential
are presented in quadrature form, using the equations (\ref{ans},
\ref{sp}) and (\ref{ww}).

\section{Classical solutions a la WKB \label{hamil}}
For our study, we shall make use of a semi-classical approximation
to extract the dynamics of the WDW equation. The semi-classical
limit of the WDW equation is achieved by taking $\rm \Psi = e^{-S}$,
and imposing the usual WKB conditions on the superpotential function
S, namely
$$\rm \left(\frac{\partial S}{\partial q} \right)^2 >> \frac{\partial^2 S}{\partial q^2}  $$

Hence, the WDW equation, under the particular factor ordering Q = 0,
becomes exactly the afore-mentioned EHJ equation (\ref{hj}) (this
approximation is equivalent to a zero quantum potential in the
Bohmian interpretation of quantum cosmology \citep{barbosa}). The
EHJ equation is also obtained if we introduce the following
transformation on the canonical momenta $\rm \Pi_q\to \partial_q S$
in Eq. (\ref{hami-i}) and then Eq. (\ref{momenta}) provides the
classical solutions of the Einstein Klein Gordon (EKG) equations.
Moreover, for  particular cases shown in table \ref{t:solutions},
the classical solutions of the EKG, in terms of $\rm q(\tau)$,
arising from Eqs. (\ref{momenta}) and (\ref{sp}) are given by
\begin{equation}
\rm gh=4\frac{d\Omega}{d\tau},\qquad
\frac{d\varphi}{\partial_\varphi Ln g}+
\frac{d\varsigma}{\partial_\varsigma Ln h}=0, \label{const-final}
\end{equation}
the second equation appears in the W function (\ref{ww2}), therefore
 W is simplified. We also have the corresponding relation with
the time $\tau$
\begin{equation}
\rm d\tau=12  \frac{1}{h} \frac{d\varphi}{\partial_\varphi
g}\,\,,\qquad d\tau=-12  \frac{1}{g}
\frac{d\varsigma}{\partial_\varsigma h} \label{time}.
\end{equation}

\subsection{Particular classical solutions}
\subsubsection{Free wave function}
 This particular case corresponds to an null potential function $U(\varphi,\varsigma)$, (see second line in table (\ref{potentials})).
 The particular exact solution for the wave function $\Psi$ becomes
\begin{equation}
\rm \Psi(\Omega,\varphi,\varsigma)=e^{ \pm \frac{s}{2}
\left(\frac{\Delta\varsigma}{\ell}+\frac{\Delta \varphi}{p}\right)}
e^{\pm \frac{1}{2}  \left(\ell \Delta \varsigma+p \Delta
\varphi\right)} e^{\frac{\Omega}{2}\left(k+Q \right)\pm
\left(3k-9\right)\frac{\Delta \varphi}{2p}}\,\, Exp\left[{-{g_0
h_0}e^{3\Omega \pm \ell \Delta \varsigma \pm p \Delta
\varphi}}\right], \label{xi}
\end{equation}

\begin{figure}[h]
\begin{center}
\includegraphics[width=15cm]{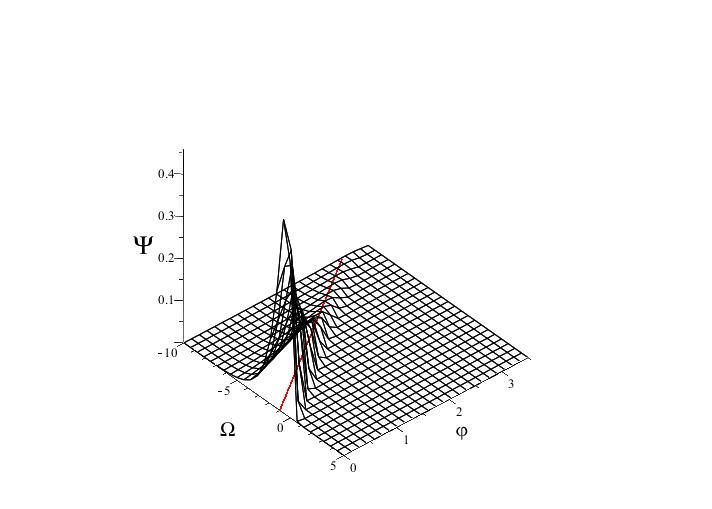}
\caption{\small{Exact wave function for the free case, i.e. for
$U(\varphi,\varsigma)=0$. The wave function (\ref{xi}) is peaked
around the classical trajectory $\Delta \Omega \pm \frac{3}{p}
\Delta \varphi=\rm \nu_0 =const$, which is the solid line shown on
the $\{\Omega,\varphi\}$ plane. For this case $\nu_0=-1$ in equation
(\ref{trajectory})} } \label{graf1}
\end{center}
\end{figure}

the classical trajectory implies that $\frac{\Delta
\varsigma}{\ell}+\frac{\Delta \varphi}{p}=0$, then there is the
simple relation between the fields $\varphi$ and $\varsigma$,
$$\rm \Delta \varsigma=-\frac{\ell}{p} \Delta \varphi.$$
So, the wavefunction can be written in terms of $\varphi$ and
$\Omega$,
\begin{equation}
\rm \Psi(\Omega,\varphi,\varsigma)= e^{\frac{k}{2}\left(\Delta
\Omega \pm \frac{3}{p}\Delta \varphi \right)+\frac{Q}{2}\Delta
\Omega}\,\, Exp\left[{-{g_0 h_0}e^{3\left(\Omega \pm
\frac{3}{p}\Delta \varphi\right)}}\right], \label{xii}
\end{equation}

Using the equantion (\ref{time}), we find the {\it classical
trajectory} on the $\{\Omega,\varphi\}$ plane as
\begin{subequations}
\label{free}
\begin{eqnarray}
\Delta \Omega \pm \frac{3}{p} \Delta \varphi&=&\rm \nu_0 =const, \label{trajectory}\\
\Delta \varphi&=& \rm Ln\left[\frac{3g_0 h_0}{4}
\Delta \tau \right]^{\pm \frac{p}{9}}, \\
\Delta \varsigma&=&\rm Ln\left[\frac{3g_0 h_0}{4}
\Delta \tau \right]^{\mp \frac{\ell}{9}},\\
\Delta \Omega&=& \rm Ln\left[\frac{3g_0 h_0}{4} \Delta \tau
\right]^{ \frac{1}{3}}+\nu_0,
\end{eqnarray}
\end{subequations}
 where the first equation corresponds to constant phase of the second exponential in
the W function (\ref{ww2}). The last equation gives the scale factor
which goes as a power law expansion, in a stiff matter epoch.

\subsubsection{ Exponential scalar potential}
For an exponential scalar potential, see the third line in table
(\ref{potentials}), the exact solution of the WDW equation and the
classical solutions are similar to the last case,  only we redefine
the constants,
$$\ell \to \sqrt{\ell^2-c_0 H_0}, \qquad p\to \sqrt{p^2+c_1G_0}.$$
We present two solutions, depending of the value of the constant
$\rm 9-c_0H_0-c_1G_0$; when $\rm 6<c_0H_0+c_1G_0<9$, the scale
factor which goes as a power law expansion, and  $\rm
c_0H_0+c_1G_0=9$, the scale factor have an exponential behavior. The
corresponding classical solutions are

\begin{enumerate}
\item{} Case $\rm 6<c_0H_0+c_1G_0<9$
\begin{subequations}
\label{free}
\begin{eqnarray}
\Delta \Omega \pm \frac{3}{\sqrt{p^2+c_1G_0}} \Delta \varphi&=&\rm \nu_1 =const, \label{trajectory2}\\
\Delta \varphi&=& \rm Ln\left[\frac{(9-c_0H_0-c_1G_0)g_0 h_0}{12}
\Delta \tau \right]^{\pm \frac{\sqrt{p^2+c_1G_0}}{9-c_0H_0-c_1G_0}}, \\
\Delta \varsigma&=&\rm Ln\left[\frac{(9-c_0H_0-c_1G_0)g_0 h_0}{12}
\Delta \tau \right]^{\mp \frac{\ell^2-c_0H_0}{9-c_0H_0-c_1G_0}},\\
\Delta \Omega&=& \rm Ln\left[\frac{(9-c_0H_0-c_1G_0)g_0 h_0}{12}
\Delta \tau \right]^{ \frac{3}{9-c_0H_0-c_1G_0}},
\end{eqnarray}
\end{subequations}
if the scale factor must have an inflationary behavior  as a power
law expansion, the exponent $\rm \frac{3}{9-c_0H_0-c_1G_0}>1$, so,
the constants $\rm 6<c_0H_0+c_1G_0<9$.
\item{}  Case $\rm c_0H_0+c_1G_0=9$

Employing the equations (\ref{const-final}) and (\ref{time}), we
choose this value before to realize the integration, obtaining
\begin{subequations}
\label{free}
\begin{eqnarray}
\Delta \Omega \pm \frac{3}{\sqrt{p^2+c_1G_0}} \Delta \varphi&=&\rm \nu_1 =const, \label{trajectory3}\\
\Delta \varphi&=& \rm \frac{g_0 h_0\sqrt{p^2+c_1G_0}}{12}\Delta
\tau, \\
\Delta \varsigma&=&\rm - \frac{g_0 h_0\sqrt{\ell^2-c_0H_0}}{12}
\Delta \tau ,\\
\Delta \Omega&=& \rm \frac{g_0 h_0}{4} \Delta \tau,\qquad
\Rightarrow \qquad A(\tau)=e^\Omega=A_0e^{\frac{g_0 h_0}{4} \Delta
\tau}.
\end{eqnarray}
\end{subequations}
\end{enumerate}
where the  scale factor have an exponential behavior, which
corresponding to inflationary epoch with exponential scalar fields.
\section{Final remarks. \label{remarks}}
Under canonical quantization we were able to determine a family of
potentials that  allows as to find exact solutions in classical and
quantum cosmology  in the inflation era. Some of the potentials of
this family satisfied the conditions that  can cross the
cosmological constant barrier $\omega_{_{DE}}=-1$,  equation
(\ref{baro}). One potential for which it to was  not possible to
solve analytically \cite{yoelsy}, by means of dynamical systems
 is possible to show that crosses
this barrier.

 The exact quantum solutions to the Wheeler-DeWitt equation were
found using the Bohmian like scheme  \citep{bohm} of quantum
mechanics, using the amplitude-real-phase approach \cite{mo-ry},
 where the ansatz to the wave function  is $\rm
\Psi(\ell^\mu) = W(\ell^\mu) e^{- S(\ell^\mu)}$ includes the
superpotential function S, which plays an important role in solving
the Hamilton-Jacobi equation.

We presented the corresponding Einstein Klein Gordon equation for
the quintom model, which is applied to the FRW cosmological model
with a barotropic perfect fluid and cosmological constant as the
matter content; the classical solutions are given in a quadrature
form for constant scalar potentials, in an inflationary stage ($p=-\rho$) these solutions are related to
the S\'aez-Ballester formalism,
\cite{saez-ballester,soco1,soco2,soco3}. For the inflationary stage with a potential that is a product of
exponentials of the fields we found two exact solutions, one with a potential expansion law and the other
with an exponential one.

\smallskip
We emphasize that the quantum potential from the Bohm like formalism
will
    work as a constraint equation which restricts our family of potentials found, see table (\ref{potentials}), \cite{sodo}.
  in this work such a problem has been solved in order
      to find the quantum potentials, which was a more important matter for being able to find the classical
      trajectories, which were showed through graphics how the classical trajectory is projected from its quantum counterpart.
      We include some steps used to solve the imaginary like
      equation (\ref{wdwho}) when we found the superpotential
      function S (\ref{sp}) and particular ansatz for the function W, we found the equation (\ref{solve}), and
      using the separation variables method we find the set of
      equations that were necessary to solve.

 The recent astronomical data suggest the
existence of the dark energy with negative pressure
\cite{perlmutter}, with ratio $\rm \omega_{_{DE}}$ between the
pressure $\rm P_{_{DE}}$ and the energy density $\rm \rho_{_{DE}}$
seems to be near or less than -1, $\rm -1.62<\omega_{_{DE}}<-0.74,$
\cite{melchiorri}. Following the equation (\ref{baro}), in order to
have agreement with these observational results, considering $\rm
V(\varsigma,\varphi)>0$ and $\rm
V(\varsigma,\varphi)>6\varphi^{\prime 2}-6\varsigma^{\prime2}$, then
the kinetic terms satisfies $\rm \varsigma^{\prime
2}<\varphi^{\prime2}$, so, the barotropic parameter $\rm
\omega_{_{DE}}<-1$. In this case, the intensity of the phantom field
$\rm \varphi$ is always larger than the intensity of the
quintessence field $\rm \varsigma$. In other case, we have that $\rm
\omega_{_{DE}}> -1$. The particular value corresponding to
cosmological constant is when the kinetic terms are null or equal in
both fields.

  However, the strange
properties of the phantom field (violation of energy conditions and
related negative energy density, the theories are not quantum
mechanically viable, either because they violate conservation of
probability, or they have unboundedly negative energy density and
lead to the absence of a stable vacuum state, \cite{Cline}). To
avoid this kind of problem, one should consider theories where the
interactions between phantom field and normal matter are as weak as
possible, but we must allow the ghosts to interact gravitationally,
since it is their gravitational interactions which are needed for
them to have any cosmological consequences. For instance, in
\cite{Nojiri1,Nojiri2}, the authors include some generalizations of
phantom cosmology with quantum contribution. Quantum effects may
lead also to negative energy density or to negative pressure what
may indicate that phantom corresponds to the effective description
of some fundamental quantum field theory (QFT). Those authors
mention that QFT can suggests some mechanism to introduce the
phantom field at the early universe, and most of energy conditions
are satisfied due to quantum effects.

\acknowledgments{ \noindent This work was partially supported by
CONACYT 179881 grant. DAIP (2011-2012) and PROMEP grants UGTO-CA-3,
UAM-I-43. PRB and MA were partially supported by UAEMex grant
FEO1/2012 103.5/12/2126. This work is part of the collaboration
within the Instituto Avanzado de Cosmolog\'{\i}a, and Red PROMEP:
Gravitation and Mathematical Physics under project {\it Quantum
aspects of gravity in cosmological models, phenomenology and
geometry of space-time}. Many calculations were done by Symbolic
Program REDUCE 3.8.
%}

%}%

\end{document}